\documentclass[12pt]{article}

\usepackage{graphicx}
\usepackage{amsmath}

\begin{document}


\begin{center}
{\Huge On the Combining Significances}
\end{center}

\begin{center}
{Sergey Bityukov, Nikolai Krasnikov, Alexander Nikitenko\\
(e-mail: Serguei.Bitioukov@cern.ch) }
\end{center}


\begin{center}
Abstract
\end{center}

We present the statistical approach to the combining of signal
significances.\\

\section{What we keep in mind as a significance ?}

The measure of the excess of observed (or expected) events in the experiment 
above the background often is named the signal significance.  
According to ref.~\cite{Frodesen} ``Common practice is to express 
the significance of an enhancement by quoting the number of 
standard deviations''. 

\bigskip

Let us distinguish the significances of two classes: 

\begin{itemize}
\item ``the initial (or internal) significance'' $S$ of 
an experiment
is the expression of two parameters of the experiment - expected number of 
signal events $N_s$ and expected number of background events $N_b$
in the given experiment (``the initial significance'' can be 
considered as a potential for discovery in planned 
experiments~\cite{BitDurh}),

\item ``the observed significance'' $\hat S$ is the expression of observed 
number of events $\hat N_{obs}$ and of the expected 
background $N_b$~\cite{Bit2000}.
\end{itemize}

The first one is a parameter of the experiment. We suppose that it is constant
for given integral luminosity.
The second one is a realization of a random variable. The observed
significance is considered as an estimator of the initial significance.

\bigskip

Why we can consider the observed significance as the realization of 
a random variable?

\bigskip

The observed number of events $\hat N_{obs}$ is the realization of the random 
variable which obeys the Poisson distribution, hence the observed 
significance $\hat S$ also is the realization of the random variable
as a function which depends from $\hat N_{obs}$.

\bigskip

It is easy to show. Let us take, as an example, the 
``counting''~\cite{GunBob} significance $\hat S_{c12}$~\cite{BitDurh}
 and the significance $\hat S_{cP}$~\cite{NarBit}. 

\noindent
The observed significance $\hat S_{c12}$  is expressed by formula 
\begin{equation}
\displaystyle \hat S_{c12} = 2 \cdot (\sqrt{\hat N_{obs}} - \sqrt{N_b}).
\end{equation}

The significance $\hat S_{cP}$ is the probability from Poisson distribution 
with mean $N_b$ to observe equal or greater than $\hat N_{obs}$ events, 
converted to equivalent number of sigmas of a Gaussian distribution, i.e.

\begin{equation}
\beta = \displaystyle 
\frac{1}{\sqrt{2\pi}}\int_{\hat S_{cP}}^{\infty}{e^{-\frac{x^2}{2}}dx},~
{\tt where}~~\beta = \displaystyle 
\sum_{i=\hat N_{obs}}^{\infty}{\frac{N_b^ie^{-N_b}}{i!}}.   
\label{eq:1}
\end{equation}

We use the method which allows to connect the magnitude of 
``the observed significance'' with the confidence density~\cite{Efr, BitSan}
of the parameter ``the initial significance''.
This method was applied in many studies~\cite{Feldman, Bit2004B}.
We carried out the uniform scanning of initial significance $S_{c12}$ and 
$S_{cP}$, varying $S_{c12}$ from value $S_{c12}=1$ 
to value $S_{c12}=16$ using step size 0.075 and varying $S_{cP}$ from 
value $S_{cP}=0$ to value $S_{cP}=6.2$ using step size 0.031. 
By playing with the two Poisson distributions
(with parameters $N_s$ and $N_b$) and using 30000 trials for each value
of $S_{c12}$ and $S_{cP}$ we used the RNPSSN function (CERNLIB~\cite{CERNLIB}) 
to construct the conditional distribution of the probability 
(the confidence density) of the production of the observed value of 
significance $\hat S_{c12}$ or $\hat S_{cP}$ by the initial significance 
$S_{c12}$ or $S_{cP}$, correspondingly. We assume that an integral 
luminosity of the experiment is a constant $N_s+N_b$. 
The parameters $N_s$ and $N_b$
are chosen in accordance  with the given initial significance $S_{c12}$ or
$S_{cP}$, the realization $\hat N_{obs}$ is a sum of realizations $\hat N_s$ 
and $\hat N_b$ of two random variables with parameters
$N_s$ and $N_b$, correspondingly.

\begin{figure}[htpb]
  \begin{center}
\includegraphics[width=0.9\textwidth]{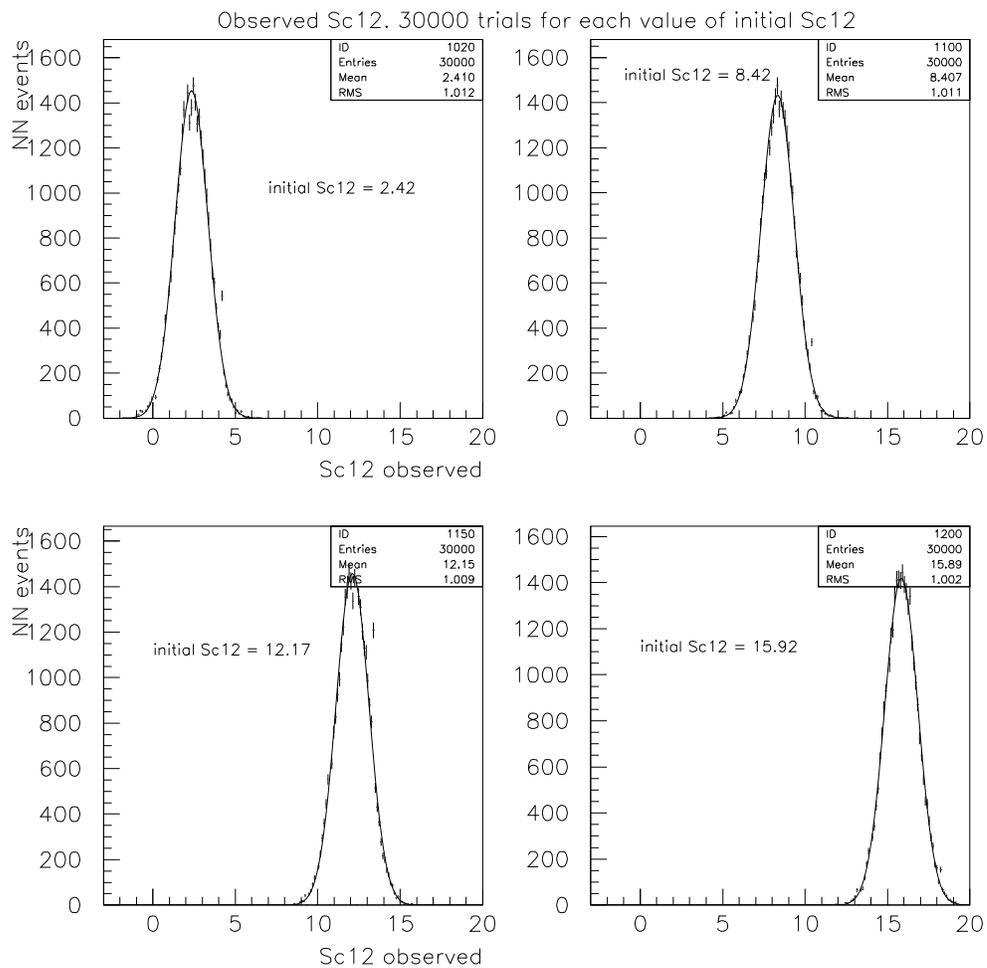} 
\caption{The observed significances $\hat S_{c12}$ for the case  
$N_s + N_b = 70$.}
    \label{fig:1} 
  \end{center}
\end{figure}

\bigskip

In Fig.1 the distributions of $\hat S_{c12}$ of several values of initial 
significance $S_{c12}$ with the given integral luminosity $N_s+N_b=70$ 
are shown. As seen, the observed distributions of significance is similar  
to the distributions of the realizations of normal distributed random 
variable with variance which close to 1. 
The distribution of the observed significance $\hat S_{c12}$ versus
the initial significance $S_{c12}$ (Fig.2) shows the result of the
full scanning.

\begin{figure}[htpb]
  \begin{center}
\includegraphics[width=0.9\textwidth]{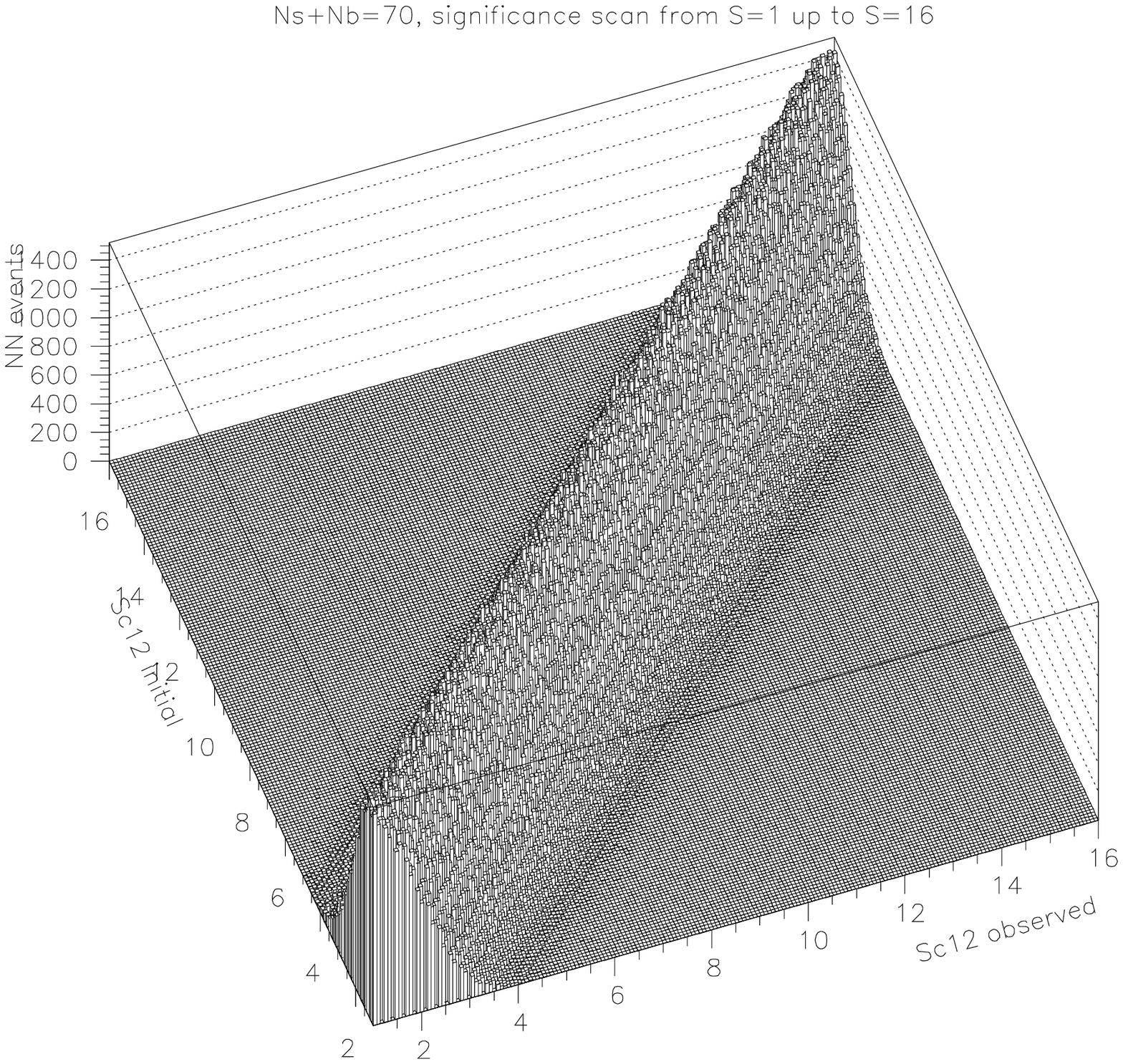} 
\caption{The distribution of observed significance $\hat S_{c12}$ versus 
the initial significance $S_{c12}$.}
    \label{fig:2} 
  \end{center}
\end{figure}

The normal distributions with a fixed variance are statistically self-dual 
distributions~\cite{BitSan}. It means that the confidence density
of the parameter ``initial significance'' $S$ has the same distribution
as the random variable which produced a realization ``the observed
significance'' $\hat S$.
The several distributions of the probability of the initial significances 
$S_{c12}$ to produce the observed values of $\hat S_{c12}$ are 
presented in Fig.3. These figures clearly shows that the observed significance
$\hat S_{c12}$ is an estimator of the initial significance 
$S_{c12}$. 

\begin{figure}[htpb]
  \begin{center}
\includegraphics[width=0.9\textwidth]{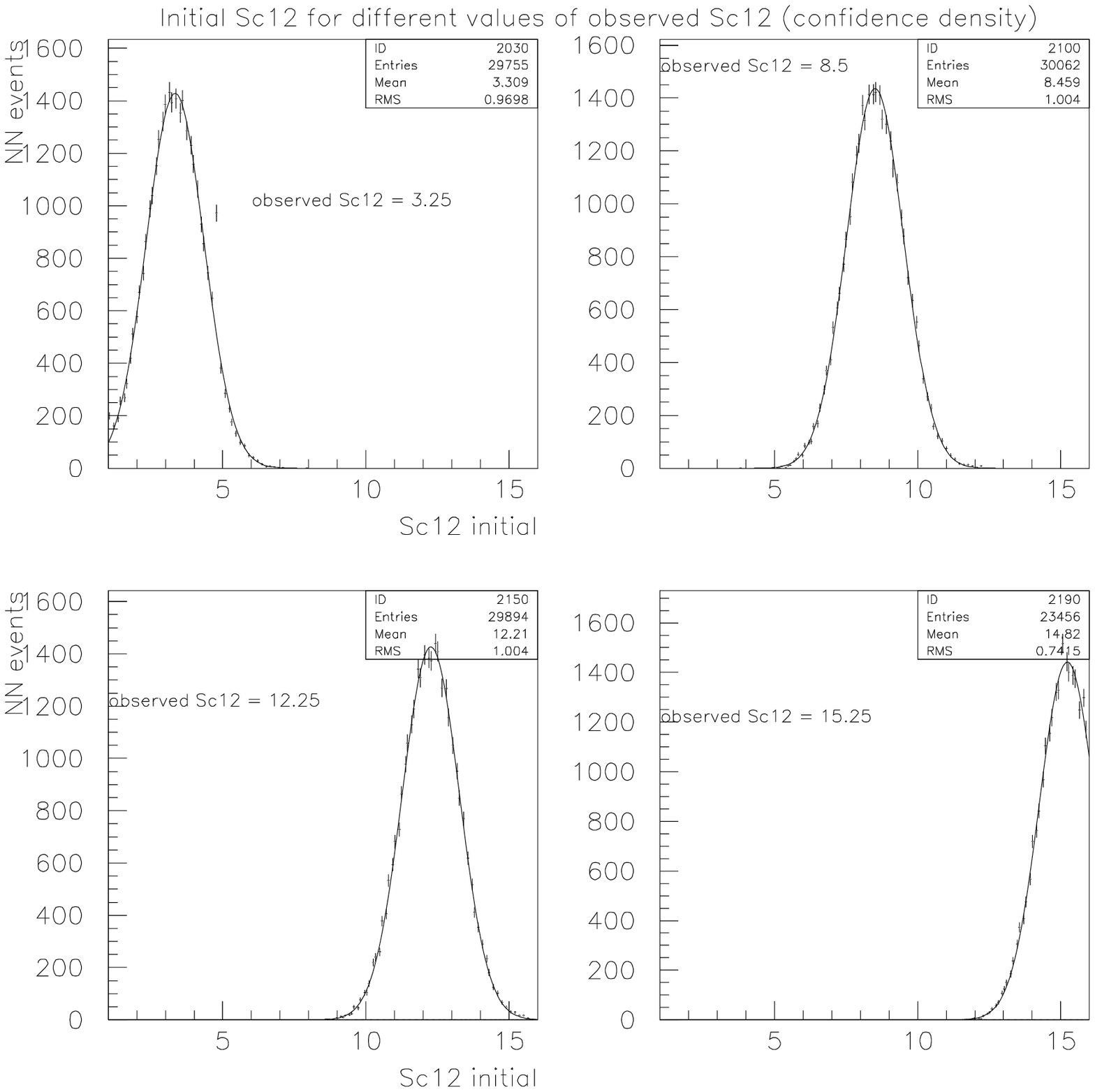} 
\caption{The distributions of the initial significances $S_{c12}$ 
(confidence densities) for the case  $N_s + N_b = 70$.}
    \label{fig:3} 
  \end{center}
\end{figure}

The distribution presented in Fig.4 shows the result of the
full scanning in the case of the observed significance $\hat S_{cP}$ and
the initial significance $S_{cP}$.

\begin{figure}[htpb]
  \begin{center}
\includegraphics[width=0.9\textwidth]{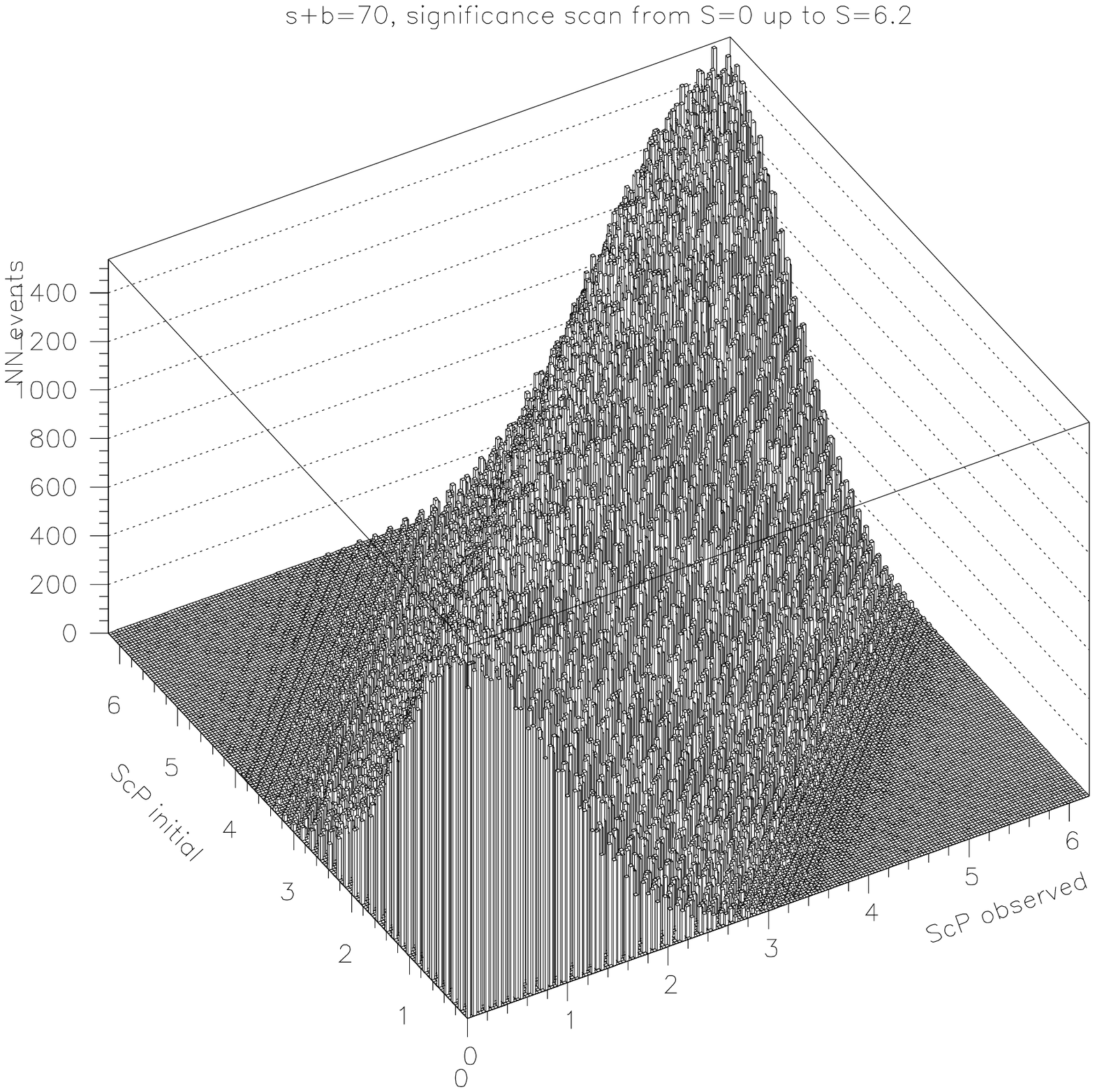} 
\caption{The distribution of observed significance $\hat S_{cP}$ versus 
the initial significance $S_{cP}$.}
    \label{fig:4} 
  \end{center}
\end{figure}

The error of these estimators with a good accuracy obeys the standard 
normal distribution (variance equals to 1). It can be confirmed by
the using of the Eqs.1-2 for pure background.
The results of the simulation of the signal absence (3000000 trials) are 
shown in Fig.5 (for the estimator $\hat S_{c12}$) and in Fig.6
(for the estimator $\hat S_{cP}$).

\begin{figure}[htpb]
  \begin{center}
\includegraphics[width=0.9\textwidth]{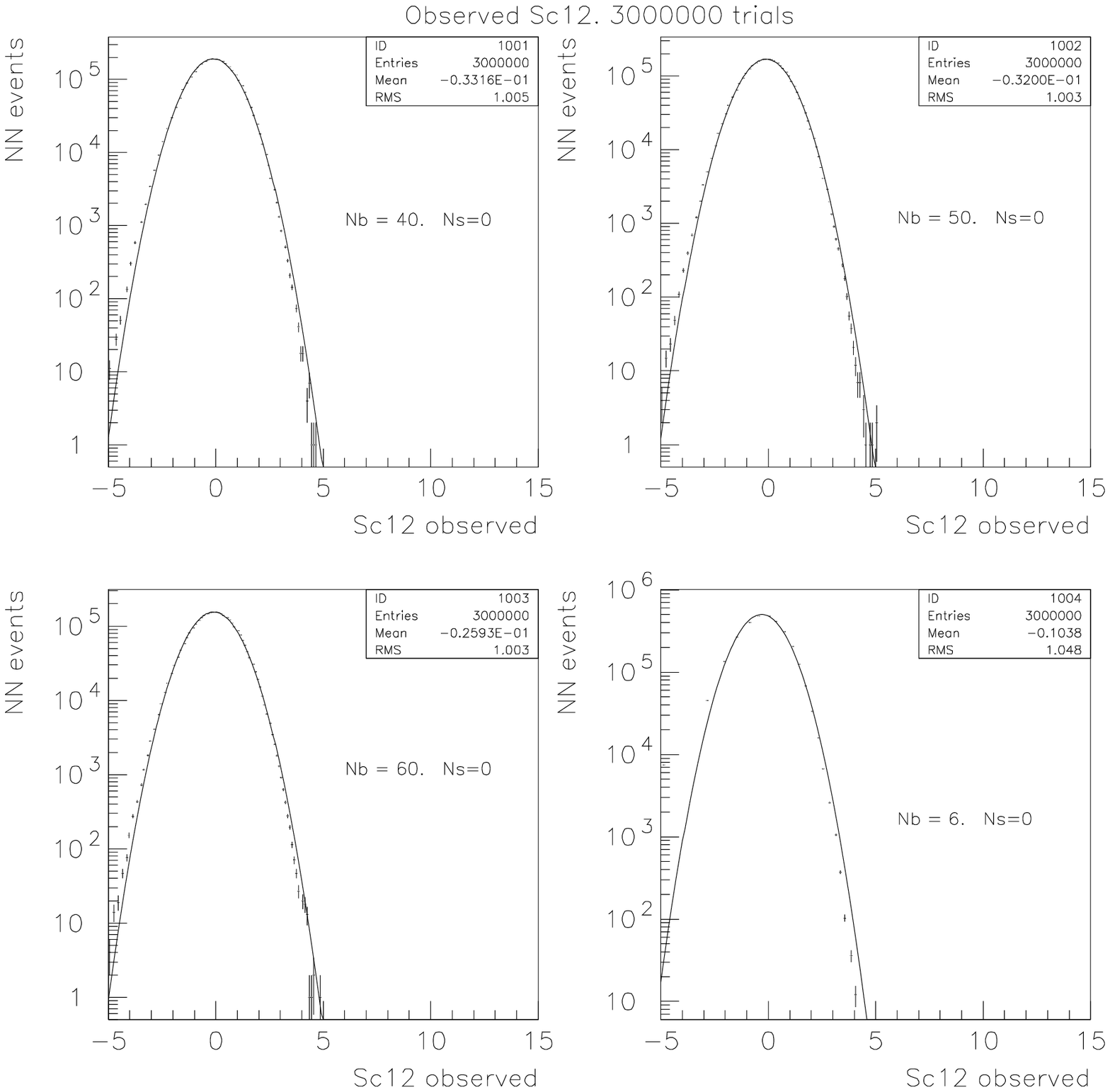} 
\caption{The distributions of the observed significances $\hat S_{c12}$ 
for four different experiments without signal.}
    \label{fig:5} 
  \end{center}
\end{figure}

\begin{figure}[htpb]
  \begin{center}
\includegraphics[width=0.9\textwidth]{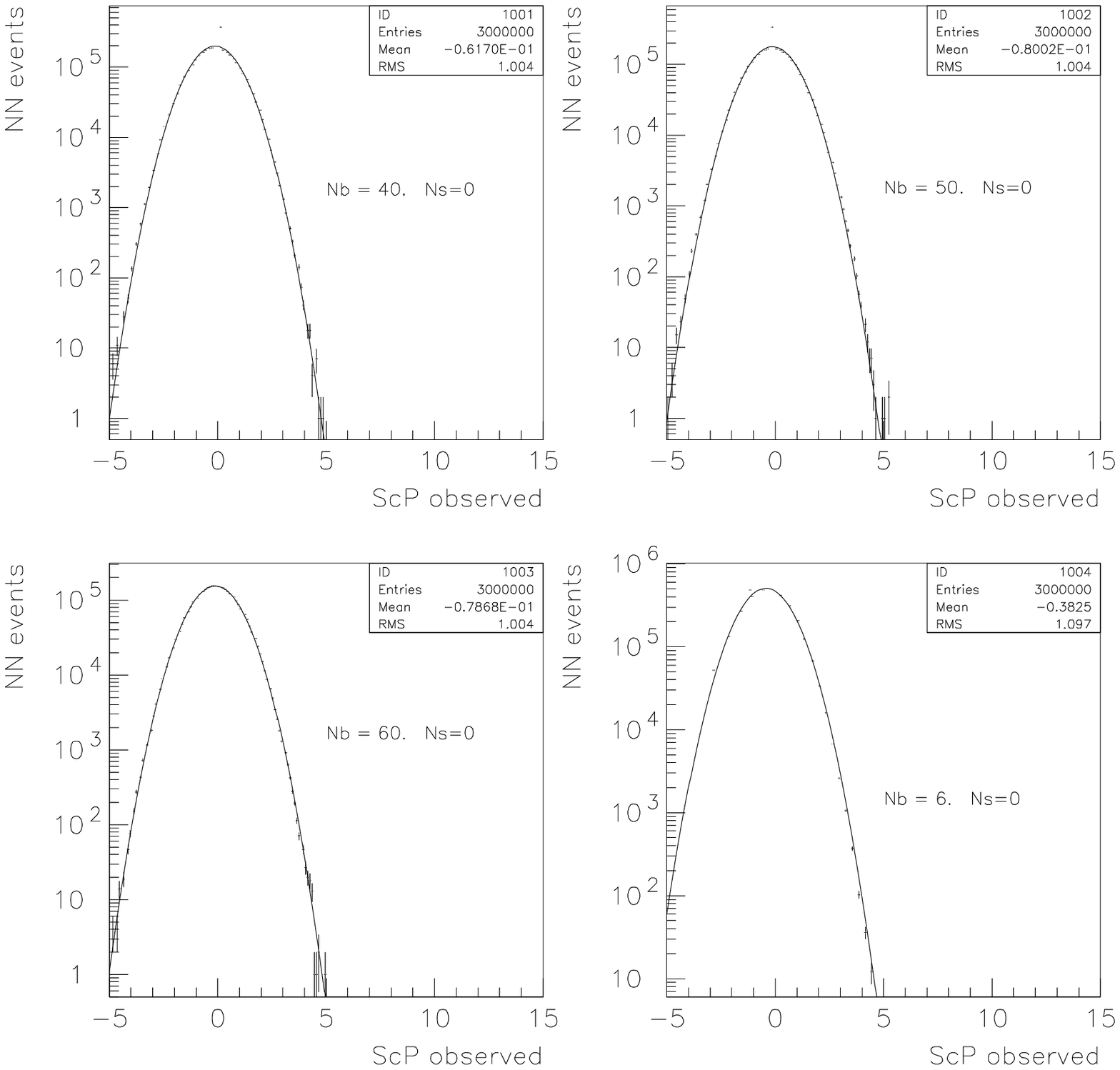} 
\caption{The distributions of the observed significances $\hat S_{cP}$ 
for four different experiments without signal.}
    \label{fig:6} 
  \end{center}
\end{figure}

\underline{
{\bf Statement 1:} The observed significance (the case of the 
Poisson flows of events)} \\
\underline{is a realization of the random variable 
which can be approximated by} \\
\underline{normal distribution with variance close to 1}.

\section{What is the Combining Significance?}

The Statement 1 allows us to determine the combinations of
the several partial significances $S_i$ as combinations of 
independent normal distributed random variables
by the simple way.

Let us define the observed sum  $\hat S_{sum}$ of partial significances and 
the observed combining significance $\hat S_{comb}$ for the $n$ observed 
partial significances $\hat S_i$ with variances $var(S_i)$:

\begin{equation} 
\displaystyle
\hat S_{sum} = \sum_{i=1}^n \hat S_i,~~~~
var(\hat S_{sum}) = \sum_{i=1}^n var(S_i),
\end{equation}

\begin{equation}
\displaystyle
\hat S_{comb} = \frac{\hat S_{sum}}{\sqrt{ var(\hat S_{sum})}}.
\end{equation}

\underline{{\bf Statement 2:} The ratio of the sum of the several
observed partial significances}\\  
\underline{and the standard deviation of this sum is 
the observed combining significance}\\ 
\underline{of several partial significances.}~
\footnote{Note the additivity of observed combined significances is not 
conserved. We must take into account the number of partial 
significances in each observed combined significance for performance 
of the Eq.3.}

\bigskip

In our case of Poisson flows of events the variances of the considered
significances close to 1. It means that the formula (Eq.4) is
approximated by the formula 

\begin{equation}
\displaystyle
\hat S_{comb} \approx \frac{\hat S_{sum}}{\sqrt{n}}.
\end{equation}

It also can be shown by Monte Carlo. Let us generate the observation of
the significances $\hat S_{c12}$ 
for four experiments with different parameters 
$N_b$ and $N_s$ simultaneously. The results of this simulation 
(30000 trials) for 
each experiment are presented in Fig.7. The distribution of the
sums of four observed significances of experiments in each trial
is shown in Fig.8 (top). Correspondingly, the Fig.8 (bottom)
presents the distribution of these sums divided by $\sqrt{4}$ in each trials,
i.e. the distribution of the observed combined significances.  

\begin{figure}[htpb]
  \begin{center}
\includegraphics[width=0.9\textwidth]{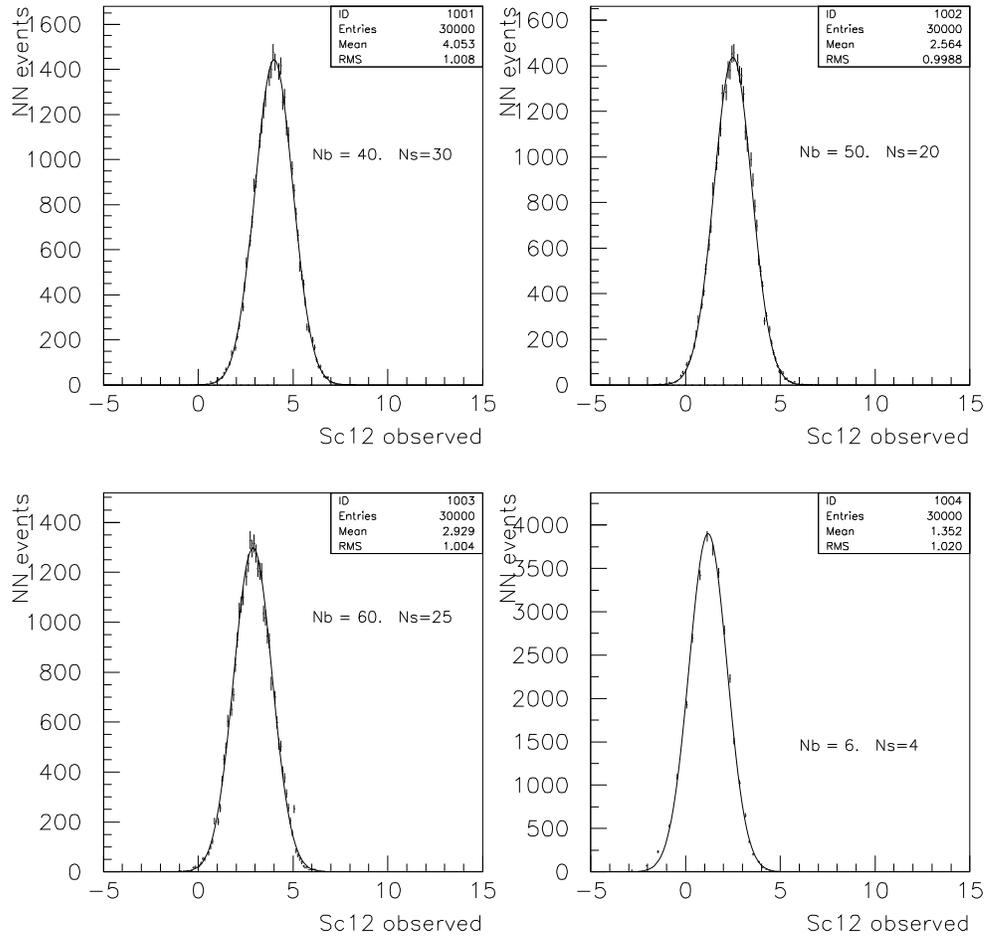} 
\caption{The distributions of the observed significances $\hat S_{c12}$ 
for four different experiments.}
    \label{fig:7} 
  \end{center}
\end{figure}

\begin{figure}[htpb]
  \begin{center}
\includegraphics[width=0.9\textwidth]{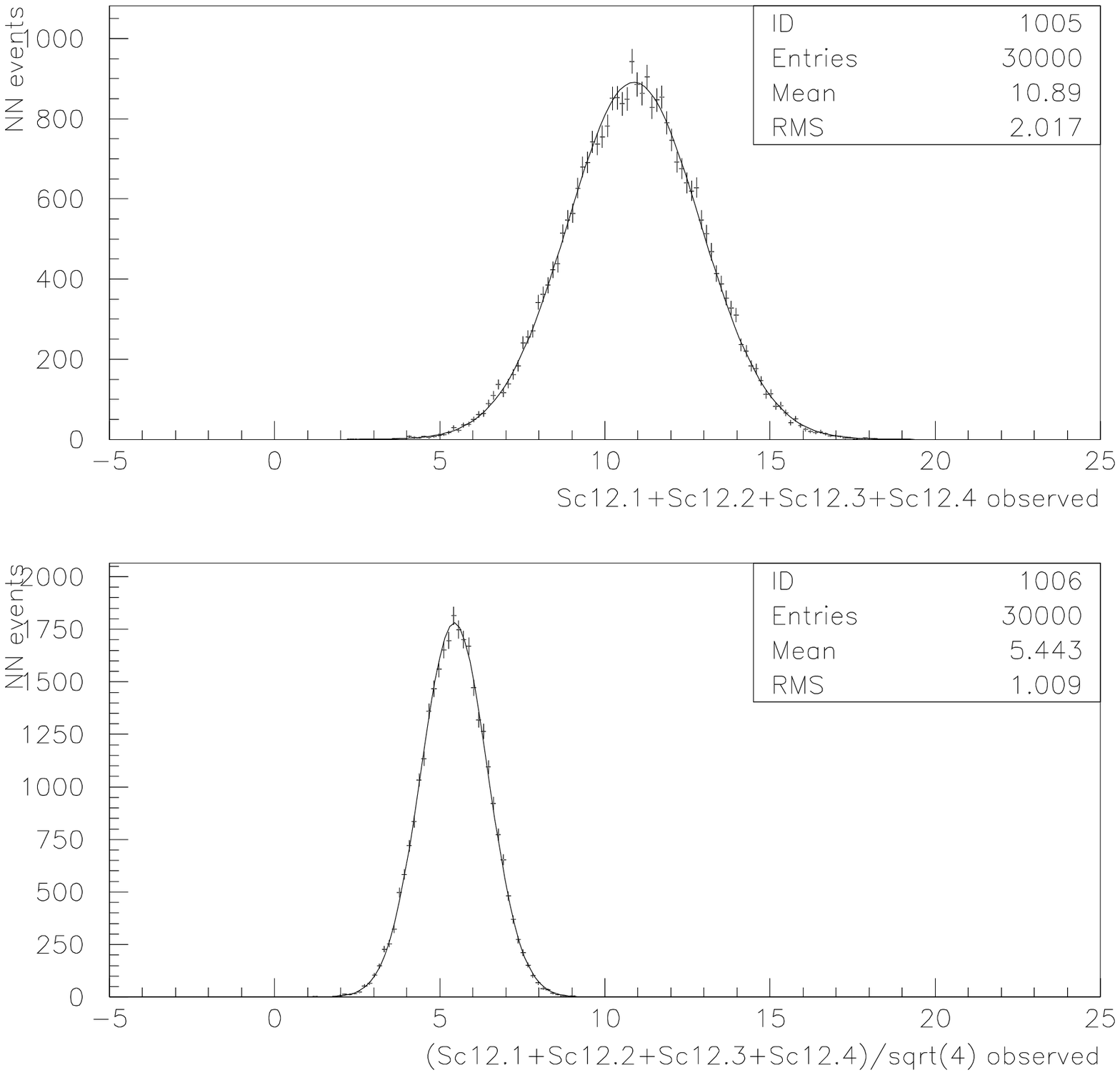} 
\caption{The distribution of the sum of observed significances 
in different experiments for each trials (top). The distribution
of the normalized sums of observed significances (bottom).}
    \label{fig:8} 
  \end{center}
\end{figure}

This property is correct also for significance $\hat S_{cP}$. 




\section{Conclusion}

The initial significance is a parameter of the given measurement.
The observed significance is a realization of the random variable.
Also the observed significance is the estimator of the
initial significance. It means that we must consider the combinations
of the significances as the combinations of the random variables with
corresponding estimators.

\section*{Acknowledgments}

We are grateful to Vladimir Gavrilov, Vassili Kachanov and 
Vladimir Obraztsov for interest and support of this work.

We also thank Alexander Lanyov, Sergey Shmatov, Vera Smirnova and 
Valeri Zhukov for very useful discussions.

This work has been partly supported by grants RFBR 05-07-90072 
and RFBR 04-02-16381.

\medskip\noindent{Key Words: }
{Uncertainty, Measurement, Estimation}

\thispagestyle{empty}
\end{document}